\documentclass[a4paper]{jpconf}
\usepackage{graphicx}
\usepackage{amsmath, amsthm, amssymb}
\newcommand{\vek}[1]{\boldsymbol{#1}}

\begin{document}
\title{Inspiral waveforms for spinning compact binaries in a new precessing convention}

\author{Anuradha Gupta$^{1}$ and Achamveedu Gopakumar$^{2}$}

\address{$^{1}$Inter University Centre for Astronomy and Astrophysics, Ganeshkhind, Pune 411007, India \\
$^2$Department of Astronomy and Astrophysics, Tata Institute of Fundamental Research, Mumbai 400005,
India
}

\ead{anuradha@iucaa.in}

\begin{abstract}
It is customary to use a precessing convention, based on
Newtonian orbital angular momentum $\vek L_{\rm N}$, to model inspiral gravitational waves 
from generic spinning compact binaries. A key feature of such a precessing convention 
is its ability to remove all spin precession induced modulations from the orbital phase 
evolution. However, this convention usually employs a post-Newtonian (PN) accurate 
precessional equation, appropriate for the PN accurate orbital angular momentum $\vek L$, 
to evolve the $\vek L_{\rm N}$-based precessing source frame. This motivated us to develop inspiral 
waveforms for spinning compact binaries in a precessing convention that explicitly use 
$\vek L$ to describe the binary orbits. Our approach introduces certain additional 3PN order 
terms in the orbital phase and frequency evolution equations with respect to 
the usual $\vek L_{\rm N}$-based implementation of the precessing convention. The implications of 
these additional terms are explored by computing the match between inspiral waveforms 
that employ $\vek L$ and $\vek L_{\rm N}$-based precessing conventions. We found that the match estimates are 
smaller than the optimal value, namely 0.97, for a non-negligible fraction of unequal mass spinning compact binaries.
\end{abstract}

\section{Introduction}
Inspiralling compact binaries containing spinning black holes (BHs) 
are plausible sources for the network of second generation gravitational wave (GW) detectors
like the advanced LIGO (aLIGO), advanced Virgo, KAGRA,
GEO-HF and the planned LIGO-India \cite{SS_lr}.
The inspiral dynamics and associated GWs from compact binaries can be accurately 
described using the post-Newtonian (PN) approximation to general relativity \cite{LR_LB}.
Moreover, an optimal detection technique of {\it matched filtering} is employed to detect and 
characterize inspiral GWs from such binaries.
In this technique, one cross correlates the interferometric output data with a bank of
templates that theoretically model inspiral GWs from spinning binaries.
The construction of these templates involves modeling the two
GW polarization states, $h_{\times}(t)$ and $h_+(t)$, associated with such events, in an accurate and efficient
manner. At present, GW frequency and associated phase evolution,  crucial inputs to compute $h_{\times, +}(t)$, are 
known to 3.5PN order
for non-spinning compact binaries \cite{BDFI} whereas the amplitudes are available to 3PN order \cite{BF3PN}.
In the case of spinning components, the spin effects enter the dynamics and GW emission via spin-orbit (SO) and spin-spin (SS) interactions \cite{BO_75}. 
Additionally, 
$\vek S_1$, $\vek S_2$ and $\vek L$, the two spin and orbital angular momenta,
 for generic spinning compact binaries
precess around the total 
angular momentum $\vek J=\vek L+\vek S_1+\vek S_2$ due to SO and SS interactions.
This forces substantial modulations of the emitted GWs from inspiralling generic spinning compact binaries  \cite{LK_95, ACST}.
Therefore, it is important to incorporate various spin effects while 
constructing inspiral GW templates for spinning compact binaries.
At present, GW frequency evolution and amplitudes of $h_{\times, +}(t)$
for maximally spinning BH binaries
are fully determined to 2.5PN and 2PN orders, respectively,  while
incorporating all the relevant spin induced effects
 \cite{LK_95, BFH}.

 There exist inspiral waveforms for precessing binaries, implemented in the {\scshape lalsimulation} package of 
LIGO Scientific Collaboration (LSC) \cite{LAL}, that employ 
the {\it precessing convention} of \cite{BCV}.
An attractive feature of this convention is its ability to remove all the spin precession induced
modulations from the orbital phase evolution. This allows one to express the orbital phase $\Phi_p(t)$
as an integral of the orbital frequency $\omega(t)$, namely $\Phi_p(t)=\int \omega(t)\, dt$.
Therefore, in this convention, the inspiral waveform from precessing binaries can be written 
as the product of a non-precessing carrier waveform and a modulation term that contains all the precessional effects.
The convention involves a {\it precessing source frame}  $(\vek e_1^l, \vek e_2^l, \vek e_3^l\equiv \hat {\vek L}_{\rm N})$
whose basis vectors satisfy the evolution equations
$\dot{\vek e}^l_{1,2, 3}=\vek \Omega_e^l \times \vek e_{1,2, 3}^l$. 
The angular frequency $\Omega_e^l$ is constructed in such a manner that the three basis vectors $\vek e_1^l, \vek e_2^l$ and $\hat {\vek L}_{\rm N}$
always form an orthonormal triad. Subsequently,  $\vek \Omega_e^l$ has to be 
$\vek \Omega_e^l= \vek \Omega_k - (\vek \Omega_k \cdot \hat{\vek L}_{\rm N})\hat{\vek L}_{\rm N}$,
where $\Omega_k$ is the usually employed precessional frequency for $\vek L_{\rm N}$. The relevant expression for $\vek \Omega_{k}$ can be obtained
by collecting the terms that multiply $\hat {\vek L}_{\rm N}$ in equation (9) in \cite{BCV}.
This triad defines an orbital phase $\Phi_p(t)$ such that  $\vek n = \cos \Phi_p \, \vek e_1^l +\sin \Phi_p \,\vek e_2^l$,
where $\vek n$ is the unit vector along binary separation vector $\vek r$. Furthermore, one can express $\dot{\vek n }$
in the co-moving frame ($\vek n, \vek \lambda = \hat{\vek L}_{\rm N} \times \vek n, \hat {\vek L}_{\rm N}$)
as $\dot{\vek n } = \dot{\Phi}_p \vek \lambda + \vek \Omega_e^l \times \vek n$.
It was argued in \cite{BCV} that $\vek \Omega_e^l$ should only be proportional to $\vek n$, leading to $\dot{\vek n } = \dot{\Phi}_p \vek \lambda$.
Thus, the adiabatic condition for the sequence of circular orbits, namely $\dot{\vek n } \cdot \vek \lambda = \omega$,
gives the desired result, i.e., $\dot{\Phi}_p=\omega$. It should be obvious that the above adiabatic condition can also imply 
$\dot{\vek n} \cdot \dot{\vek n} =  \omega^2$.

In practice, the precessional equation for $\vek L$ is employed to construct $\vek \Omega_e^l$ and to evolve $\vek L_{\rm N}$.
As a consequence, $\vek \Omega_e^l$ is no longer proportional to $\vek n$ \cite{GG1}
and this leads to PN corrections to  $\dot {\Phi}_{p} = \omega$ (see section 2.1 of \cite{GG4} for detailed calculation). 
These observations motivated us to provide a set of PN accurate equations to obtain temporally evolving 
quadrupolar order $h_{\times, +}$ for generic spinning compact binaries
in an $\vek L $-based precessing convention. 
In the next section, we present our $\vek L$-based precessing convention and explore its data analysis implications in the later section.

\section{Inspiral waveforms via an $\vek L$-based precessing convention}
\label{Sec_L}
 
In this section, we introduce a $\vek k$-based precessing source frame 
($\vek e_1$,$\vek e_2$, $\vek e_3 \equiv \vek k$), to develop a $\vek k$-based precessing convention, where $\vek k$ is unit vector along $\vek L$.
The precessional dynamics of $\vek e_1, \vek e_2$ and $\vek e_3$ are provided by 
$\dot {\vek e}_{1,2,3}=\vek \Omega_e \times \vek e_{1,2,3}$, where 
$ \vek \Omega_e \equiv \vek \Omega_k -  ( \vek \Omega_k \cdot \vek k)\, \vek k$ and
$ \Omega_k$ is the usual precessional frequency of $\vek k$.
It should be obvious that 
$\dot {\vek e}_{3} = \vek \Omega_e \times \vek e_{3} $  is identical to
$\dot {\vek e}_{3} =  \vek \Omega_k \times \vek e_{3}$ 
as $ \vek e_{3} \equiv \vek k$.
It is possible to
construct a $\vek k$-based 
 co-moving triad ($\vek n, \vek \xi=\vek k \times \vek n, \vek k$) and define an
orbital phase $\Phi$ such that $\vek n = \cos \Phi \, \vek e_1 + \sin \Phi \, \vek e_2$ and $\vek \xi = - \sin \Phi \, \vek e_1 + \cos \Phi \, \vek e_2$.
Also, the time derivatives of $\vek n$ is given by $\dot { \vek n}  = \dot {\Phi}\, \vek \xi + \vek \Omega_e \times \vek n$.
Consequently, the frame independent adiabatic condition for circular orbits, namely
$ \dot { \vek n} \cdot \dot { \vek n} \equiv \omega^2$, leads to $\omega^2 = \dot{\Phi}^2 + \Omega_{e\xi}^2$, 
where $\Omega_{e\xi} =\vek \Omega_e \cdot \vek \xi$. 
This results  in the following 3PN accurate differential equation for $\Phi$,
\begin{equation}
\label{Eq_phidot}
 \dot {\Phi} = \frac{c^3}{G\, m}\, x^{3/2} \, \biggl\{  1- \frac{x^3}{2}\, \Bigl[\delta_1 \, q\, \chi_1 \, (\vek s_1\cdot \vek \xi) 
 + \frac{\delta_2}{q}\, \chi_2\, (\vek s_2\cdot \vek \xi)\Bigr]^2\biggr\} \,,
\end{equation}
where $q=m_1/m_2$, $\delta_{1,2} = \eta/2 + 3\,(1\mp \sqrt{1-4\eta})/4$,  
 $\eta=m_1\,m_2/m^2$ and $m=m_1+m_2$. The PN expansion parameter $x$ is defined as $(G\, m \,\omega/c^3)^{2/3}$. 
The Kerr parameters $\chi_1$ and $\chi_2$ of the two compact objects of mass $m_1$ and $m_2$ specify 
their spin angular momenta by $\vek S_{1,2}=G\, m_{1,2}^2\, \chi_{1,2}\,\vek s_{1,2}/c$, 
where $\vek s_1$ and $\vek s_2$ are the unit vectors along $\vek S_1$ and $\vek S_2$.
The use of $\vek L$ to describe binary orbits also modifies the evolution 
equation for $\omega$ (or $x$). This is because  
the SO interactions are usually incorporated in terms of $\vek s_1 \cdot \vek l$ and $\vek s_2 \cdot \vek l$,
in the literature \cite{LK_95,BCV}.
These terms require modifications due to the 1.5PN order relation between $\vek l$ and $\vek k$ which can be obtained from equation~(8) in \cite{GG4}. 
The PN accurate expression for $\dot{x}$ along with the 3PN additional terms is given by equation (9) in \cite{GG4}.
Notice that these additional terms are, for example, with respect to equation (3.16) in \cite{Bohe2013}
that provides PN accurate expression for $\dot{x}$ while invoking $\vek l$ to describe
binary orbits. 

We now model inspiral GWs from spinning binaries in our 
$\vek k$-based precessing convention. 
The expressions for quadrupolar order $h_{\times}$ and $h_+$, written in the
frame-less convention \cite{LK_95}, read
\begin{subequations}
\label{Eq_hp_hx}
\begin{eqnarray}
 h_{\times}|_{\rm Q}(t) &=&  2\, \frac{ G\, m\, \eta \, x}{c^2\, R'}\, (2\, \xi_x\, \xi_y -2\, n_x\, n_y)\,, \\
 h_{+}|_{\rm Q}(t) &=&   2\, \frac{ G\, m\, \eta \, x}{c^2\, R'}\, (\xi_x^2 - \xi_y^2- n_x^2 + n_y^2)     \,,
\end{eqnarray}
\end{subequations}
where $\xi_{x,y}$ and $n_{x, y}$ are the $x$ and $y$ components of $\vek \xi$ and $\vek n$ in an inertial frame
associated with  $\vek N$,
the unit vector that points from the source to the detector,  while $R'$ is the  distance to the binary.
These  $x$ and $y$ components of $\vek \xi$ and $\vek n$ can be expressed in terms of the  Cartesian 
components of $\vek e_1$ and $\vek e_2$. 
In order to obtain $h_{\times}|_{\rm Q}(t)$ and $h_{+}|_{\rm Q}(t)$, we require to solve numerically the differential equations for 
$\Phi, x, \vek e_1$ and $\vek e_2$.
We use equation~(\ref{Eq_phidot}) for $\Phi$ while the differential equation for $x$, $\vek e_1$  and $\vek e_2$ are given by equations~(9) 
and (13) in \cite{GG4}.
It easy to see that the evolution of $\vek e_1$ and $\vek e_2$ depends upon the time variation of $\vek s_1$, $\vek s_2$ and $\vek k$.
Therefore, we also need to solve differential equations for $\vek s_1$, $\vek s_2$ and $\vek k$. These differential equations that include the leading order SO
and SS interactions can be obtained from equation~(15) in \cite{GG4}.

In practice, we 
numerically solve 
the differential equations for $\vek e_1, \vek k$, $\vek s_1$, $\vek s_2$, $\Phi$ and $x$ to
obtain temporally evolving Cartesian components of $\vek \xi$ and $\vek n$.
Note that we 
 do not solve the differential equation for $\vek e_2$. This is because the temporal evolution of 
 $\vek e_2$ can be estimated using the relation $\vek e_2(t)=\vek k(t)\times \vek e_1(t)$.
The required initial values for the Cartesian components of $\vek e_1$, $\vek k$, $\vek s_1$ and $\vek s_2$ are given by freely 
choosing the following five angles: $\theta_{10}$, $\phi_{10}$, $\theta_{20}$, $\phi_{20}$ and $\iota_0$.
The initial Cartesian components of $\vek s_1$, $\vek s_2$,  $\vek k$ and $\vek e_1$ as functions of the above angles are given by equations~(16) in \cite{GG4}. 
Note that this choice of initial conditions is influenced by the {\scshape lalsuite} SpinTaylorT4 code of LSC.
Additionally, we let the initial $x$ value to be  $x_0=(G\, m\, \omega_0/c^3)^{2/3}$ where $\omega_0=10\pi$ Hz (relevant for aLIGO)
and the initial phase $\Phi_0$ to be zero. In what follows, we explore the data analysis implications of these inspiral
waveforms that employ the $\vek L$-based precessing convention.

\section{ Implications of inspiral waveforms in $\vek L$-based precessing convention}
\label{result}
 We employ the {\it match} \cite{DIS98} to compare inspiral waveforms constructed via
the $\vek l$ and $\vek k$-based precessing conventions.
Our comparison is influenced (and justified) by the fact that the precessing source frames of these
two conventions are functionally identical.
This should be evident from the use of {\it the same} precessional frequency, appropriate
for $\vek k$, to obtain PN accurate expressions for both the $\vek l$-based $\vek \Omega_e^l$ and 
$\vek k$-based $\vek \Omega_e^l$.
Therefore, the {\it match} estimates probe influences of the additional 3PN order
terms present in the differential equations for $\Phi$ and $x$ in our approach.
Note that these 3PN order terms are not present in 
the usual implementation of the precessing
convention as provided by the {\scshape lalsuite} SpinTaylorT4 code.

 Our match ${\cal M}(h_l, h_k)$ computations involve $h_l $ and $ h_k$, the two families of inspiral
waveforms arising from the $\vek l$ and $\vek k$-based precessing conventions.
The $ h_l$ inspiral waveform families are adapted from the {\scshape lalsuite} SpinTaylorT4 code of LSC
while $ h_k$ families arise from our approach (equations~(\ref{Eq_hp_hx})).
We employ  the quadrupolar order expressions
for $h_{\times, +}$ while computing $h_l$ and  $h_k$ in the present analysis.
Moreover, the two families are characterized by 
identical values of $m, \eta, \chi_1 $ and $ \chi_2$. Also, the initial
orientations of the two spins in the $\vek N$-based inertial frame are also chosen to be identical.
The computation of $\vek N \cdot \vek l$ from $\vek N \cdot \vek k$ with the help of  equation~(8) in \cite{GG4}
ensures that $\vek l$ and $\vek k$  orientations  at the initial epoch are physically equivalent.
Therefore, our  match computations indeed compare two waveform families with physically 
equivalent orbital and spin configurations at the initial epoch.
Note that we 
terminate  $h_l $ and $ h_k$ inspiral waveform families when their respective
$x$ parameters reach $0.1$ ($r\sim 10 \, G\, m/c^2 $).

\begin{figure}[h]
\includegraphics[width=14pc]{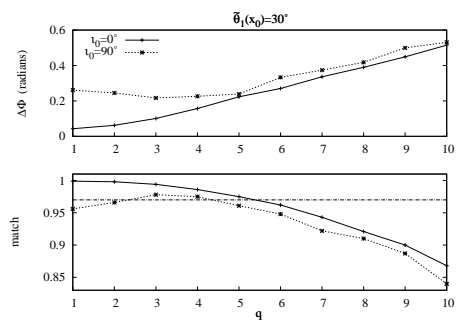}\hspace{2pc}%
\begin{minipage}[b]{16pc}\caption{\label{figure:q_M_Phi}Plots for the  accumulated orbital phase ($\Delta
\Phi$) and the associated match ($\mathcal{M}$) estimates
as functions of the mass ratio $q$ for maximally spinning  $m=30M_{\odot}$
BH binaries inspiralling in the $[x_0, 0.1]$ frequency interval.}
\end{minipage}
\end{figure}

Figure~\ref{figure:q_M_Phi} represents the result of our ${\cal M}$ computations.
The binary configurations have initial dominant SO misalignments 
$\tilde\theta_1(x_0)$ $(\cos^{-1}(\vek k \cdot \vek s_1))$ as $30^{\circ}$ and we
let the initial orbital plane orientation in the $\vek N$-based inertial frame to take 
two values leading to edge-on ($\iota_0= 90^{\circ}$) and face-on ($\iota_0= 0^{\circ}$) binary orientations.
For these two configurations, we need to choose $\theta_{10}$ to be $30^{\circ}$ and $60^{\circ}$, respectively. Moreover, we choose $\phi_{10}=0^{\circ}$,
$\theta_{20}=20^{\circ}$, $\phi_{20}=90^{\circ}$.
Let us note that $\vek l$ orientations (from $\vek N$) for these configurations will be slightly 
different from $0^{\circ}$ or $90^{\circ}$ due to the 1.5PN accurate relation between $\vek l$ and 
 $\vek k$.
 
We also plot $\Delta \Phi$, the accumulated orbital phase differences in
the frequency interval $[x_0, 0.1]$. 
We find that the variations in ${\cal M}$ estimates are quite independent of the initial 
orbital plane orientations. 
We see a gradual decrease in  ${\cal M}$ values as we increase the $q$ value
and this variation is reflected in the gradual increase of $\Delta \Phi$.
Incidentally, this pattern is also observed for configurations having somewhat smaller initial dominant SO misalignments.
However, the ${\cal M}$ estimates are close to unity for tiny $\tilde \theta_1(x_0)$ values 
and this is  expected as precessional effects 
are minimal for such binaries.
Therefore, the effect of the above discussed additional 3PN order terms are more pronounced for 
high mass ratio compact binaries having {\it moderate} dominant SO misalignments.

We find that the match estimates are less than the optimal 0.97 value for a non-negligible fraction 
of unequal mass spinning compact binaries. It may be recalled that 
such an optimal match value roughly corresponds to a $10\%$ loss in the ideal event rate.
We, therefore, conclude that the additional 3PN order terms in frequency and phase evolution equations in our approach should not be neglected
for a substantial fraction of unequal mass binaries.



\section*{References}
\begin{thebibliography}{9}
\bibitem{SS_lr}
 Sathyaprakash B S and Schutz B F 2009 {\em Living Rev. Relativity} {\bf 12}

   
 
 \bibitem{LR_LB}
  Blanchet L 2006 {\em Living Rev. Relativity} {\bf 9} 4
  

\bibitem{BDFI}
Blanchet L, Damour T, Esposito-Farese G and Iyer B R 2004  {\em Phys. Rev. Lett.} {\bf 93} 091101 


\bibitem{BF3PN}
 Blanchet L, Faye G, Iyer B R and Sinha S, 2008 {\em Class. Quantum Grav.} {\bf 25} 165003 

 
 
 \bibitem{BO_75}
  Barker B and O'Connell R 1975 {\em Phys. Rev. D} {\bf 12} 329 
  
  \bibitem{LK_95}
   Kidder L 1995 {\em Phys. Rev. D} {\bf 52} 821
 
 \bibitem{ACST} 
   Apostolatos T A,  Cutler C,  Sussman G J and Thorne K 1994 {\em Phys. Rev. D} {\bf 49} 6274
 
  
  \bibitem{BFH}
  Buonanno A, Faye G and Hinderer T 2013 {\em Phys. Rev. D} {\bf 87} 044009; references therein
  
  
  \bibitem{LAL} http://www.ligo.org/index.php
     
\bibitem{BCV}
  Buonanno A,  Chen Y and Vallisneri M 2003 {\em Phys. Rev. D} {\bf 67} 104025 
    
  
\bibitem{GG1}
Gupta A and Gopakumar A 2014 {\em Class. Quantum Grav.} {\bf 31} 065014, arXiv:1308.1315

\bibitem{GG4} Gupta A and Gopakumar A 2015 {\em Class. Quantum Grav.} {\bf 32} 175002, arXiv:1507.00406.


  \bibitem{Bohe2013} 
  Bohe A, Marsat S and Blanchet L 2013 {\em Class. Quantum Grav.} {\bf 30} 135009   
         
  \bibitem{DIS98}
Damour T, Iyer B R and Sathyaprakash B S 1998 {\em Phys. Rev. D} {\bf 57} 885 

    
\end{thebibliography}
\end{document}